\documentclass[A4, 11pt]{article}
\usepackage{amsmath}
\usepackage{amssymb}

\begin{document}

\title{On the Security of ``an efficient and complete remote user authentication scheme''}

\author{Manik Lal Das\\Dhirubhai Ambani Institute of Information and Communication Technology\\Gandhinagar - 382007, India.\\
Email: maniklal\_das@daiict.ac.in}

\date{}
\maketitle

\begin{abstract}
Recently, Liaw et al. proposed a remote user authentication scheme
using smart cards. Their scheme has claimed a number of features
e.g. mutual authentication, no clock synchronization, no verifier
table, flexible user password change, etc. We show that Liaw et
al.'s scheme is completely insecure. By intercepting a valid login
message in Liaw et al.'s scheme, any unregistered user or
adversary can easily login to the remote system and establish a
session key.\vspace{2 mm}
\end{abstract}

\section{Introduction}
Remote system authentication is a process by which a remote system
gains confidence about the identity (or login request) of the
communicating partner. Since the Lamport's scheme \cite{lam81},
several remote user authentication schemes and improvements have
been proposed with and without smart cards. Recently, Liaw et al.
\cite{lia06} proposed a remote user authentication scheme using
smart cards. Their scheme has claimed a number of features e.g.
mutual authentication, no clock synchronization, no verifier
table, flexible user password change, etc. We show that Liaw et
al.'s scheme is completely insecure. Any unregistered user can
easily login to the remote system and establish a session key.

\section{The Liaw et al.'s scheme}
The scheme consists of five phases: registration, login,
verification, session and password change.\\
\textit{Registration phase}: A new user $U_i$ submits identity
$ID_i$ and password $PW_i$ to the remote system for registration.
The remote system computes $U_i$'s secret information $v_i =
h(ID_i, x)$ and $e_i = v_i \oplus PW_i$, where $x$ is a secret key
maintained by the remote system and $h(\cdot)$ is a secure one-way
hash function. Then the remote system writes $h(\cdot)$ and $e_i$
into the memory of a smart card and issues the card to $U_i$.\\
\textit{Login phase}: When $U_i$ wants to log into the remote
system, he/she inserts the smart card into the terminal and enters
$ID_i$ and $PW_i$. The smart card then performs the following
operations:
\newcounter{1}
\begin{list}{L\arabic{1}.}
{\usecounter{1}} \item Generate a random nonce $N_i$ and compute
$C_i = h(e_i \oplus PW_i, N_i)$. \item Send the login message $<
ID_i, C_i, N_i >$ to the remote system.
\end{list}

\textit{Verification phase}: To check the authenticity of $< ID_i,
C_i, N_i >$, the remote system checks the validity of $ID_i$. If
$ID_i$ is valid, computes $v_i^{\prime} = h(ID_i, x)$ and checks
whether $C_i = h(v_i^{\prime}, N_i)$. Then generates a random
nonce $N_s$, encrypts the message $M = E_{v_i^{\prime}}(N_i, N_s)$
and sends it back to the card.\\
The smart card decrypts the message $D_{e_i \oplus PW_i}(M)$ and
gets $(N_i^{\prime}, N_s^{\prime})$. Then verifies whether
$N_i^{\prime} = N_i$ and $N_s^{\prime} = N_s$\footnote{It is noted
that the verification of $N_s^{\prime} = N_s$ cannot be examined
because the smart card does not have information about $N_s$}. If
these checks hold valid, the mutual authentication is done.\\
\textit{Session phase}: This phase involves two public parameters
$q$ and $\alpha$ where $q$ is a large prime number and $\alpha$ is
a primitive element mod $q$. The phase works as follows:
\newcounter{5}
\begin{list}{S\arabic{5}.}
{\usecounter{5}}\item The remote system computes $S_i =
\alpha^{N_s}$ mod $q$ and sends $S_i$ to the smart card. The smart
card computes $W_i = \alpha^{N_i}$ mod $q$ and sends $W_i$ to the
remote system. \item The remote system computes $K_s =
(W_i)^{N_s}$ mod $q$ and, the smart card computes $K_u =
(S_i)^{N_i}$ mod $q$. It is easy to see that $K_s = K_u$. Then,
the card and the remote system exchange the data using the session
key and $e_i$.
\end{list}

\textit{Password change phase}: With this phase $U_i$ can change
his/her $PW_i$ by the following steps:
\newcounter{6}
\begin{list}{S\arabic{6}.}
{\usecounter{6}}\item Calculate $e_i^{\prime} = e_i \oplus PW_i
\oplus PW_i^{\prime}$. \item Update $e_i$ on the memory of smart
card to set $e_i^{\prime}$.
\end{list}

\section{Security Weaknesses}
\textit{Weakness of Authentication phase}: The authentication
phase suffers from the replay attacks. The authenticity of the
login request is not checked at all. The adversary $\mathcal{A}$
(or any unregistered user) intercepts a valid login request, say
$< ID_i, C_i, N_i>$. Later $\mathcal{A}$ sends $< ID_i, C_i, N_i
>$ to the remote system, as a login request . To validate $< ID_i,
C_i, N_i >$, the remote system does the following:
\newcounter{11}
\begin{list}{\arabic{11}.}
{\usecounter{11}} \item Check the validity of $ID_i$. This holds
true, because the adversary sends $ID_i$, intercepted from a valid
login request. \item Compute $v_i^{\prime} = h(ID_i, x)$ and check
whether $C_i = h(v_i^{\prime}, N_i)$. This check also passes
successfully, because there is no record at the server side
whether $N_i$ was used in some previous login message. Therefore
the server is unable to detect whether the $C_i$ is coming from a
legitimate user or from an adversary. Now we see the security
strength of the mutual authentication. \item The remote system
generates a nonce $N_s^*$ and encrypts the message $M =
E_{v_i^{\prime}}(N_i, N_s^*)$, then sends $< M >$ back to the
communicating party (assumes logged in entity is a legitimate
user). \item $\mathcal{A}$ will not do anything, simply sends a
valid signal by saying that the server authenticity is done and
then, $\mathcal{A}$ gains the access to the remote system.
Therefore, ultimately there is no user or server authenticity
checks at all.
\end{list}
\textit{Weakness of Session phase}: Although Liaw et al.'s scheme
used Diffie-Hellman \cite{dif76} key exchange protocol for session
key establishment; however, they did not consider the risk of
Diffie-Hellman's protocol (i.e., man-in-the-middle attack) while
establishing the user and server common session key. Let us
examine the weakness of the session phase.
\newcounter{15}
\begin{list}{\arabic{15}.}
{\usecounter{15}}\item The remote system computes $S_i =
\alpha^{N_s^*}$ mod $q$ and sends $S_i$ to the communicating
party. $\mathcal{A}$ (who already passes the authentication phase
and gains the access to the remote system) computes $W_i =
\alpha^{N_i}$ mod $q$ and sends $W_i$ to the remote system. \item
The remote system computes $K_s = (W_i)^{N_s^*}$ mod $q$ and
$\mathcal{A}$ computes $K_a = (S_i)^{N_i}$ mod $q$. It is easy to
see that $K_s = K_a$.
\end{list}
In fact, all the parameters $N_i, S_i, W_i, \alpha, q$ are public,
thereby any one can compute the session key. Once the session key
is established then the remote system and $\mathcal{A}$ exchange
data in an encrypted manner, where $e_i$ acts as the encryption
key. Firstly, the remote system does not know $e_i$. Secondly, the
session key never serve the purpose of the transaction privacy,
instead it is just xor-ed with the message and $e_i$ is used for
transaction privacy, which is not the actual scenario in the
practical applications.\vspace{2mm}\\ \textit{Weakness of Password
change phase}: There is no verification of the entered password.
This effectively makes the smart card useless. Suppose $U_i$
enters his password which is actually misspelled or incorrect,
that is, instead of $PW_i$ he/she enters $PW$. However, the smart
card takes the wrong password $PW$ and asks for a new password.
Now, $U_i$ enters new password $PW_i^{\prime}$. The smart card
updates old $e_i$ by the new $e_i^{\prime}$ where $e_i^{\prime} =
e_i \oplus PW \oplus PW_i^{\prime} = h(ID_i, x) \oplus PW_i \oplus
PW \oplus PW_i^{\prime}$. In the next login time, $U_i$ cannot
login to the remote system, because the verification of $C_i$
fails. In another scenario, if $U_i$'s smart card is lost or
stolen, then the party who got the smart card, would try to login
and enters some random password, which leads to block the card, as
there is no provision of checking the entered password.

\section{Conclusion}
We have shown the security weaknesses of the Liaw et al.'s scheme.
The design of the Liaw et al.'s scheme is so weak that any one can
login to the remote system by just intercepting a valid login
message.

\end{document}